\documentclass[conference]{IEEEtran}
\IEEEoverridecommandlockouts

\usepackage{fancyhdr}
\fancypagestyle{TitlePage}{%
   \fancyhf{}%
    \fancyhead[L]{\vspace{0.1cm} \small \bf Preprint, In: Proceedings of the 2026 IEEE 103rd Vehicular Technology Conference (VTC2026-Spring), Nice, France}
    \fancyfoot[C]{\small \bf \thepage} 
}

\usepackage{microtype}
\usepackage{amsmath,amssymb,amsfonts}
\usepackage{textcomp}
\usepackage{xcolor}
\usepackage{graphicx}
\usepackage[utf8]{inputenc}
\usepackage{ragged2e}
\usepackage[T1]{fontenc}
\usepackage[numbers]{natbib}
\usepackage{textcomp}
\usepackage{units}
\usepackage{bm}
\usepackage{url}
\usepackage{eurosym}
\usepackage{txfonts}

\usepackage{scalerel}
\usepackage{tikz}
\usetikzlibrary{svg.path}

\definecolor{orcidlogocol}{HTML}{A6CE39}
\tikzset{
  orcidlogo/.pic={
    \fill[orcidlogocol] svg{M256,128c0,70.7-57.3,128-128,128C57.3,256,0,198.7,0,128C0,57.3,57.3,0,128,0C198.7,0,256,57.3,256,128z};
    \fill[white] svg{M86.3,186.2H70.9V79.1h15.4v48.4V186.2z}
                 svg{M108.9,79.1h41.6c39.6,0,57,28.3,57,53.6c0,27.5-21.5,53.6-56.8,53.6h-41.8V79.1z M124.3,172.4h24.5c34.9,0,42.9-26.5,42.9-39.7c0-21.5-13.7-39.7-43.7-39.7h-23.7V172.4z}
                 svg{M88.7,56.8c0,5.5-4.5,10.1-10.1,10.1c-5.6,0-10.1-4.6-10.1-10.1c0-5.6,4.5-10.1,10.1-10.1C84.2,46.7,88.7,51.3,88.7,56.8z};
  }
}
\newcommand\orcidicon[1]{\href{https://orcid.org/#1}{\mbox{\scalerel*{
\begin{tikzpicture}[yscale=-1,transform shape]
\pic{orcidlogo};
\end{tikzpicture}
}{|}}}}

\graphicspath{{graphics/}}
\usepackage{fouridx}
\usepackage{xfp}
\graphicspath{{figures/}}

\newcommand{\mrm}[1]{{\mathrm{#1}}}

\newcommand{\LBP}{{$L_\mrm{BP}$}}

\def\BibTeX{{\rm B\kern-.05em{\sc i\kern-.025em b}\kern-.08em
    T\kern-.1667em\lower.7ex\hbox{E}\kern-.125emX}}

\usepackage[hidelinks,linkcolor=black,citecolor=black]{hyperref} %

\definecolor{delim}{RGB}{0,0,0}
\definecolor{numb}{RGB}{106, 109, 32}
\definecolor{string}{rgb}{0,0.0,0.0}

\begin{document}

\title{
Experimental Evaluation of LPWAN Technologies: mioty, LoRaWAN, Sigfox, NB-IoT, and LTE-M\\ in Deep Indoor Environments 
}

\newcommand{\wisun}{{Wi-SUN\textsuperscript{\tiny \textregistered}}}
\newcommand{\lora}{{LoRa\textsuperscript{\tiny \textregistered}}}
\newcommand{\loraalliance}{{LoRa Alliance\textsuperscript{\tiny \textregistered}}}
\newcommand{\lorawan}{{LoRaWAN\textsuperscript{\tiny \textregistered}}}
\newcommand{\mioty}{{mioty\textsuperscript{\tiny \textregistered}}}
\newcommand{\miotyalliance}{{Lo Alliance\textsuperscript{\tiny \textregistered}}}

\author{\IEEEauthorblockN{Christof Röhrig, Benz Cramer} 
\IEEEauthorblockA{\textit{IDiAL -- Institute for the Digital Transformation of Application and Living Domains} \\
\textit{Dortmund University of Applied Sciences and Arts}\\
Dortmund, Germany \\
christof.roehrig@fh-dortmund.de,  \orcidicon{0000-0002-3286-3703} 0000-0002-3286-3703} 
}

\maketitle
\thispagestyle{TitlePage}
\pagestyle{fancy}

\begin{abstract}
Low Power Wide Area Networks (LPWAN) are often used in applications such as Smart City, Smart Buildings and Smart Metering.
Energy meters are often located in underground spaces that are difficult to reach with wireless technology.
This paper presents an experimental study comparing different LPWAN technologies in terms of building penetration.
The technologies \mioty, Low Power Long Range Wide Area Network (\lorawan), Sigfox, Narrow Band Internet of Things (NB-IoT), and Long Term Evolution for Machines (LTE-M) are evaluated experimentally.
The aim of the research is to investigate the performance of building penetration of different radio technologies in real-life scenarios.
The measurements are performed with inexpensive off-the-shelf equipment.
The results of the study are integrated in the metering system of the Dortmund University of Applied Sciences and Arts. 

\end{abstract}


\maketitle

\section{Introduction}
The Dortmund University of Applied Sciences and Arts operates several university buildings spread across the city of Dortmund.
Energy meters are located in underground rooms and energy supply tunnels that are difficult to reach with wireless technology.
The paper presents an experimental study that compares different LPWAN (Low Power Wide Area Network) technologies in terms of building penetration and radio coverage.
These LPWAN technologies differ in terms of energy consumption, building penetration, radio coverage and radio range.
The experiments will be carried out in various university buildings and in an underground parking garage with six levels half high. %

This paper extends the work we have presented in \cite{roehrig:cobee25} and presents the following contribution:
We evaluate five LPWAN technologies in demanding environments in terms of building penetration and radio coverage. 
The technologies \mioty, Low Power Long Range Wide Area Networks (\lorawan), Sigfox, Narrow Band Internet of Things (NB-IoT), and Long Term Evolution for Machines (LTE-M) are evaluated experimentally. 
In \cite{roehrig:cobee25} we have compared \lorawan{} with \wisun, Sigfox and NB-IoT. 
As \mioty{} becomes more widespread and offers greater reliability compared to \lorawan, \mioty{} is being experimentally compared with \lorawan.
In a second environment we compare three LPWAN technologies (Sigfox, NB-IoT, LTE-M) operated by network operators with each other. %

\section{Problem definition and methodical approach}
This research compares five LPWAN radio technologies for smart metering in deep indoor and underground environments.
The aim of the research is to investigate the performance of building penetration of different radio technologies in real-life scenarios.
The measurements are performed with inexpensive off-the-shelf equipment. We focus on measuring the building penetration loss (BPL) without considering the power consumption of the equipment.
The BPL $L_\mrm{BP}$ is a logarithmic attenuation measure that is specified in \unit{dB}. 
 
We consider two scenarios: In the first scenario, the radio infrastructure, such as gateways (GWs) and base stations (BSs), is installed inside the building. 
Radio waves propagate through walls and ceilings. %

According to \cite{ITU:25}, the theoretic path-loss model for indoor environments has the form %
\begin{equation}
L_\mathrm{indoor} = L(d_0) + N \cdot \log_{10} \left(\frac{d}{d_0} \right) + L_f(n),
\label{eq:indoor}
\end{equation}
where $N$ is the distance power loss coefficient, $d$ is the distance between the BS and the device, $d_0$ is the reference distance,
$L(d_0)$ is the path loss at $d_0$, $L_f$ is the floor penetration loss, $n$ is the number of floors between BS and device ($L_f =$ 0~dB for $n = 0$).

In the second scenario, the radio infrastructure (GWs, BSs) is installed in the outdoor environment. 
The radio waves propagate through the windows or the exterior wall into the building (Outdoor to Indoor, O2I). 
The penetration loss O2I depends heavily on the material, which the radio waves have to penetrate. %
The path loss incorporating O2I can be modeled %
\begin{equation}
L_\mrm{O2I} = L_\mrm{b} + L_\mrm{tw} + L_\mrm{indoor} + \mathcal{N}(0,\sigma_\mrm{P}^2),
\label{eq:o2i}
\end{equation}
where $L_\mrm{b}$ is the basic outdoor path loss,  $L_\mrm{tw}$ is the path loss through the external wall, 
$L_\mrm{indoor}$ is the inside loss dependent on the depth into the building,
and $\sigma_\mrm{P}$ is the standard deviation for the penetration loss.
Values for $L_\mrm{tw}$ are given in \cite{3GPP:25}, e.g. $L_\mrm{concrete}/\mrm{dB} = 5 + 4 \cdot f / \mrm{GHz}$ and $L_\mrm{glass}/\mrm{dB} = 2 + 0.2 \cdot f / \mrm{GHz}$.

Due to the frequency dependence of the BPL, low frequencies are preferred when good building penetration is required.
An additional loss $L_\mrm{npi}$ must be added if the radio signal propagates through the external wall by non-perpendicular incidence \cite{3GPP:25}.

It is very difficult to calculate the BPL based only on theoretical models because radio waves propagate in different ways in the building due to reflections and diffraction. 
In deep indoor and underground scenarios, such as basements and tunnels, the real signal path, and therefore the entire path loss, is too complex to be accurately represented by a linear model \cite{Malarski:19}.
Therefore, we chose an empirical experimental evaluation where the received signal strength indicator (RSSI) is measured at outdoor measurement points (MPs) at ground level and then compared with the RSSI measured at indoor MPs on different floors and underground MPs in basements and (see Fig.~\ref{fig:EFS}).
We measure the BPL 
\begin{equation}
\label{eq:lbp}
L_\mrm{BP} = P_\mrm{outdoor} - P_\mrm{indoor}
\end{equation}
as the difference between the received signal strength outdoors $P_\mrm{outdoor}$ and the received signal strength indoors $P_\mrm{indoor}$, both measured in \unit{dBm}.
Aim of the study is to compare off-the-shelf devices in a real-world environment.

\section{Related Work}
LPWAN technologies have received increased attention in recent years. 
They are characterized by their long range and low power consumption. 
Orlovs et al. provide a literature survey of LPWAN technologies in real-world deployments \cite{orlovs:2025}.
LPWAN technologies known for their good building penetration include \mioty, \lorawan, Sigfox, NB-IoT, LTE-M. 

\mioty{} is a relatively young technology, which is why there are only a few experimental studies on \mioty{} in real environments in the published literature.
Robert and Lauterbach compare \mioty{} with \lorawan{} with laboratory measurements in \cite{robert26}.
Zerai et. al. present a comparative study for \lorawan{} and \mioty{} in an industrial environment \cite{Zerai23}.
Joubert presents an experimental study comparing \mioty{} with \lorawan{} at Schneider Electric in Grenoble \cite{joubert:25}.
Schreiber and Fosalau provide a theoretical comparision of \mioty{} with NB-IoT with focus on metering applications \cite{schreiber:sielmen25}.
Sikora et al. compare \mioty, \lorawan{}, Sigfox and NB-IoT theoretically and provide an experimental evaluation in a unified testbed \cite{sikora:indin19}. 
Oberacher et. al. report experiences from a large-scale \mioty{} installation \cite{oberascher:2024}.

Kadusic et al. provide an overview of \lorawan, NB-IoT and Sigfox and the characteristics of these radio technologies \cite{Kadusic:22}. 
Naumann and Oelers compare the energy consumption and the link budget of NB-IoT, \lorawan{} and Sigfox based on theoretical models \cite{Naumann:21}.
Trendov et. al. present a comparative study for \lorawan{}, NB-IoT, Sigfox and LTE-M in urban outdoor environments (cities of Halle and Köthen) \cite{trendov:gepecom25}.
Stusek et al. provides a comparative study of LPWAN propagation models in urban scenarios \cite{stusek:20}. They compare propagation models for \lorawan{}, Sigfox, and NB-IoT.
Roosipuu et al. investigate the use of NB-IoT for monitoring and control of smart urban drainage systems. Different depths of devices in manholes are investigated \cite{Roosipuu:23}.  
Thrane et al. present an experimental evaluation of an NB-IoT propagation model for deep indoor scenarios. The measurement campaign was conducted in a system of long underground tunnels \cite{Thrane:20}.

\section{Technology Overview}
LPWAN technologies known for good building penetration operate in the sub-GHz spectrum and include \mioty, \lorawan, Sigfox, NB-IoT, and LTE-M. 
A large maximum coupling loss (MCL) is a prerequisite for good building penetration.
The MCL describes the maximum difference between the transmitted and received power at the sender and receiver, respectively.

\mioty{}, \lorawan and Sigfox operate in the unlicensed spectrum of \unit[868]{MHz} in Europe, 
while NB-IoT and LTE-M operate in the licensed spectrum of \unit[800]{MHz} (band 20) or \unit[900]{MHz} (band 8) in Europe. 
To ensure fair use of the unlicensed spectrum, each device is limited in transmit power and duty cycle (DC). 
In the \unit[868]{MHz} band, the transmit power is normally limited to \unit[14]{dBm} (\unit[25]{mW}) and a DC of 0.1\%, 
except in the G1 band (868.0 - \unit[868.6]{MHz}) where a DC of 1\% is allowed and
the G3 band (869.4 - \unit[869.65]{MHz}) where a transmit power of \unit[27]{dBm} (\unit[500]{mW}) and a DC of 10\% is allowed.
The G3 band is typically used for downlink (DL, from BS to device) messages where a BS is connected to many devices and transmits at high power. 
In contrast, quality of service can be guaranteed in the licensed spectrum, where the mobile network operator (MNO) controls the use of the spectrum and the DC is not limited. 

\mioty{} is an LPWAN protocol that uses telegram splitting (TS) and ultra narrow-band (UNB) in the unlicensed spectrum. 
TS divides a data telegram into several sub-packets and sends them after applying forward error correction (FEC) in a partially predefined time and frequency pattern.
This makes the transmission robust against interferences and packet collisions. Thanks to the FEC, up to 50\% of sub-packets can be compensated for.  
Each sub-packet requires a signal bandwidth (BW) of \unit[2.3]{kHz} and an air time of \unit[15.14]{ms} in normal mode (EU1).  
TS is generated in a software-defined radio (SDR) and uses standard Gaussian Minimum Shift Keying (GMSK) modulation, which can be produced using standard sub-GHz radio chips.
The standard is defined in the specification TS 103 357 of the European Telecommunications Standards Institute (ETSI). 
In Europe, \mioty{} is using the G1 band for the uplink (UL, from device to BS) and the G3 band for the DL.
It requires a BW of \unit[200]{kHz} for two channels in UL and DL (EU1).
The data rate of \mioty{} is \unitfrac[0.5]{kbit}{s}, the payload of a telegram can be up to 250 bytes.
The strengths of \mioty{} are its low power consumption of \unit[17.8]{\textmu Wh} (device, EU1) per message and 
its ability to support an ultra-high device density with up to 3.5 million messages per day and BS \cite{schreiber:sielmen25}.
Further details on the physical layer of \mioty{} can be found in \cite{robert26}.

\lorawan{} is based on the proprietary \lora{} radio technology developed by Scemtech.  
\lorawan{} defines the communication protocol and system architecture and is managed by the open \loraalliance. 
It can achieve data rates between \unitfrac[0.25]{kbit}{s} and \unitfrac[11]{kbit}{s} in Europe.
It uses spreading factors (SF) to adjust the data rate and receiver sensitivity. A higher SF increases sensitivity, but reduces data rate.
The payload of the messages can be from 51 to 222 bytes in Europe depending on the data rate. The number of messages is limited by the DC regulations in Europe and may be even lower due to restrictions by the \lorawan{} network operator. 
The Things Network's (TTN) fair use policy limits a device's airtime to \unit[30]{s} per day. \lorawan{} networks can be operated by a network operator, a community such as TTN, or self-deployed.  
\lorawan{} thus offers a high degree of flexibility for the end user.

The proprietary Sigfox radio technology was developed by the French company Sigfox S.A. (now owned by UnaBiz). 
It uses UNB with a signal BW of \unit[0.1]{kHz} and achieves long range and requires low power. 
The Sigfox network is based on a star topology and requires a local Sigfox network operator to carry the traffic generated.
In Europe, Sigfox is using the G1 band for the UL and the G3 band for the DL.
Each message is transmitted by default three times on different, randomly selected frequencies. 
This channel accessing scheme is named Random Frequency Division Multiple Access (RFDMA).
The demodulation in the BS is done by SDR which analyzes the total band, to detect transmitted signals and to retrieve sent data.
Sigfox supports up to 140 UL and 4 DL messages per day, each carrying a payload of 12 bytes and 8 bytes respectively, at a data rate of \unitfrac[0.1]{kbit}{s}. %
Sigfox subscription cost is 10\,€  per year for one device and 140 UL, 4 DL messages per day (Heliot Europe). 

NB-IoT operates in licensed spectrum and is standardized by the 3rd Generation Partnership Project (3GPP) as LTE Cat-NB1 and -NB2.
It is only available through MNOs. Implementing NB-IoT is cost effective as it is based on existing cellular infrastructure.
Compared to other LPWAN technologies that operate in unlicensed spectrum, the transmit power and DC are not limited by regulation.
Therefore, the MCL is higher, resulting in better building penetration. The downside is higher power consumption.
NB-IoT offers data rates of up to \unitfrac[62]{kbit}{s} in Cat NB1 and up to \unitfrac[159]{kbit}{s} in Cat NB2 in the UL 
and up to \unitfrac[26]{kbit}{s} and \unitfrac[127]{kbit}{s} respectively in the DL.  
Applications are typically low throughput, delay tolerant and low mobility. Examples are smart meters and remote sensors.
The cost of the subscription is 11\,€ for a SIM card, \unit[500]{MByte} for 10 years (1NCE, IoT Lifetime Flatrate).

LTE-M operates in licensed spectrum and is standardized by the 3GPP as LTE Cat-M1 (3GPP release 13) and Cat-M2 (3GPP release 14).
Compared to NB-IoT the data rate is large in UL and DL: \unitfrac[1]{Mbit}{s} for Cat-M1 and 4 Mbit/s - \unitfrac[7]{Mbit}{s} for Cat-M2.
The latency is much lower compared to NB-IoT: 10 – 15 ms for Cat-M1 and 50 – 100 ms Cat-M2.  
The downside is the lower receiver sensitivity and therefore a lower MCL.
The costs of subscription are the same as for NB-IoT at the same data volume (1NCE).

Table~\ref{tab:lpwan} compares the key parameters of the technologies (at highest sensitivity and MCL). 
The MCL is the maximum possible path loss from transmitter to receiver. A high value is important for penetrating underground environments. 
The values are taken from \cite{Naumann:21} and the data sheets of the experimental equipment in Table~\ref{tab:dev}.
\begin{table}[h!]
\caption{Comparison of different LPWAN technologies (Europe)}
\label{tab:lpwan}
\begin{center}
{\small
\begin{tabular}{|l|c|c|c|c|}
\hline
Technology        & \mioty{}      & \lora{}       & Sigfox  & Cat-NB / M   \\\hline
Band MHz          &  868          & 868           & 868     &  800 / 900   \\\hline
Tx pow. GW$^2$    & 27            & 14/27$^1$     & 27      & 23           \\\hline
Tx pow. dev.$^2$  & 14            & 14            & 14      & 23           \\\hline
Rx sens. GW$^2$   & -135$^5$      & -140          & -142    & -141/-131    \\\hline 
Rx sens. dev.$^2$ & -128          & -136          & -126    & -139/-129    \\\hline
MCL UL$^3$        & 149           & 154           & 156     & 164/154      \\\hline
MCL DL$^3$        & 155           & 150/163$^1$   & 153     & 162/152      \\\hline
BW$^4$            & 2.3           & 125           & 0.1     & 3.75/15      \\\hline
price module      & 15€           & 6€            & 8€      & 7€/14€      \\\hline
annual cost       & -             & -             & 10€     & 1-2€         \\\hline
Dev. availab.     & 0             & ++            & -       & +            \\
\hline
\end{tabular}
}
\end{center}
{\tiny \textbf{1}: depending on the frequency of use,~~ \textbf{2}: in dBm,~~ \textbf{3}: in dB,~~ \textbf{4}: in kHz,~~\textbf{5}: depending on the SDR frontend}
\end{table}

The sensitivity of a receiver to a particular LPWAN technology is specified at a given signal-to-noise ratio (SNR). 
The SNR limit depends on the modulation scheme of the technology.
The \lora{} receiver sensitivity is $\unit[-140]{dBm}$ at the GW with an SNR of $\unit[-20]{dBm}$.
The Sigfox receiver is sensitive to signals with an SNR of $\unit[-142]{dBm}$ at the BS with an SNR of $\unit[9]{dBm}$.
The SNR values can not be compared directly because the BW of the technologies differs large. \lorawan{} uses a BW $B=\unit[125]{kHz}$ and Sigfox $B=\unit[0.1]{kHz}$.
For thermal noise density $N_0=\unitfrac[-174]{dBm}{Hz}$, the thermal noise level is $\unit[-123]{dBm}$ at $B=\unit[125]{kHz}$ and $\unit[-154]{dBm}$ at $B=\unit[0.1]{kHz}$.
Therefore, at the same noise density, Sigfox's noise power is $\unit[31]{dB}$ lower than LoRa's.
Due to the different BWs of the various LPWAN technologies, the SNR is not compared experimentally.

\section{Experimental evaluation of LPWAN technologies}
\subsection{Experimental setup}
We use several off-the-shelf devices for the experiments (see Table~\ref{tab:dev}). 
This equipment does not provide an accurate measurement of RSSI, but is intended to provide an estimate of building penetration using off-the-shelf devices.  
\begin{table*}[h!]
\caption{Devices used in experiments}
\label{tab:dev}
\begin{center}
{\small
\begin{tabular}{| l | c | c | c | c | c | c | c |}
\hline
Device              & AVA     & MUNIA-M   & M3B Magnolinq & LPS8   & LoRa E5 mini  & MKR FOX   & SIM7080G         \\\hline %
type                & GW      & device    & device        & GW      & device       & device    & device           \\\hline
manufacturer        & Weptech & Weptech   &  LZE GmbH     & Dragino & Seeed        & Arduino   & Simcom           \\\hline
radio chip          &  SDR    & TI CC1310 & TI CC1310     & SX1308  & STM32WLE     & ATAB8520E & MDM9205          \\\hline %
technology          & \mioty  & \mioty    & \mioty        & LoRa    & LoRa         & Sigfox    & NB-IoT / LTE-M   \\\hline
band MHz            & 868     & 868       &  868          & 868     & 868          & 868       &  900 / 800       \\\hline
Tx pow. / dBm       & 14      & 14        & 14            & 14 / 20 & 14           & 13        & 20               \\\hline
Rx sen. / dBm       & -135    & -128      & -128          & -140    & -136.5       & -126      & -130 / -107       \\\hline 
\end{tabular}
}
\end{center}
\end{table*}
The receiver sensitivities in Table~\ref{tab:dev} are taken from the data sheets and represent the best-case scenario for low noise. 
In a noisy environment, however, the sensitivity of the receiver can be much lower than that published in the data sheets.  
The \mioty{} AVA GW provides an estimate of the receiver sensititvity based on noise measurements. 
In our experimental environment, the sensitivity of the GW is estimated to be $\unit[-125]{dBm}$, based on a measured noise density of $\unitfrac[-156]{dBm}{Hz}$. 
The sensitivity undergoes a change over time due to its dependence on the fluctuating noise power.

The experiments were conducted in two different environments. 
As there is poor Sigfox coverage on the university campus (see \cite{roehrig:cobee25}), 
for the first measurements an environment was chosen in which Sigfox, NB-IoT and LTE-M could be received with a strong signal.
For \mioty{} and \lorawan, an infrastructure with dedicated GWs was set up on the campus.

\subsection{O2I penetration of Sigfox, NB-IoT and LTE-M (public infrastructure)}
The first measurement campaign was conducted in an underground car park with six half underground levels (A-F) in the centre of Dortmund.
In this environment, the Sigfox, NB-IoT and LTE-M BSs are close to the car park, resulting in high outdoor RSSI values.
The distance between the experimental environment and the NB-IoT/LTE-M BS is approximately \unit[100]{m}. 
The location of the Sigfox base station is unknown.
Since the RSSI level received from the Sigfox BS is relatively high (outdoors: $\unit[-61]{dBm}$), the BS is likely not far away. 
The signal received from the next public \lorawan{} GW (distance ~ \unit[800]{m}) is weak (outdoors: $\unit[-96]{dBm}$), therefore \lorawan{} is not included in this experimantal study. 

We utilize commercial LPWAN networks. %
For Sigfox Heliot provides the network, for NB-IoT and LTE-M, Dt. Telekom is choosen.
Measurements are made on different underground floors to measure the penetration loss of the ceilings.
The signal strength values for NB-IoT and LTE-M are measured DL and are obtained directly from the modem using AT commands. 
The device SIM7080G provides several signal strength and quality parameters. In this research, the Reference Signal Received Power (RSRP) is choosen,
because it provides more accurate signal power estimations by excluding interference from other sectors.  
The Sigfox RSSI values are measured by the BS in UL direction and reported by the Sigfox back-end system.

Table~\ref{tab:garage} shows the results of the experimental evaluation. The RSSI values for level 0 are measured outdoors at the entry of the car park and
are the reference values for calculating \LBP{} using (\ref{eq:lbp}).
In this measurement campaign, NB-IoT provides better building penetration than LTE-M and Sigfox, which cant't reach the lowest level F.
\LBP{} is significantly higher for Sigfox compared to NB-IoT and LTE-M. 
This could be caused by a flatter angle of entry of the signal into the car park, which would result in a higher attenuation of the signal.

\begin{table}[!ht]
\caption{RSSI values in $\unit{dBm}$ and \LBP{} in $\unit{dB}$ for Sigfox, NB-IoT and LTE-M}
\label{tab:garage}
    \centering
    \begin{tabular}{|l|l|r|r|r|}
    \hline
        level & & Sigfox & NB-IoT & LTE-M  \\ \hline \hline
        O & RSSI & -61 & -53 & -60 \\ \hline
        A & RSSI & -103 & -84 & -87 \\ \hline
        ~ & \LBP & 42 & 31 & 27 \\ \hline
        C & RSSI & -113 & -95 & -98 \\ \hline
        ~ & \LBP & 52 & 42 & 38 \\ \hline
        E & RSSI & -134 & -112 & -112 \\ \hline
        ~ & \LBP & 73 & 59 & 52 \\ \hline
        F & RSSI & - & -123 & - \\ \hline
        ~ & \LBP & - & 70 & - \\ \hline
    \end{tabular}
\end{table}

\begin{table}[b!]
\caption{RSSI values in $\unit{dBm}$, \LBP{} in $\unit{dB}$ and PER in \% for \mioty{} and \lorawan{}}
\label{tab:RSSI}
\begin{center}
{\small
\begin{tabular}{|c|c|c|c|c|c|c|}
\hline
              & \multicolumn{3}{|c|}{\mioty} & \multicolumn{3}{|c|}{\lorawan}\\ \hline
           MP & RSSI & PER & $L_\mrm{BP}$ & RSSI & PER & $L_\mrm{BP}$\\ \hline
            I & -42  & 0   & -            & -34  & 0   & - \\ \hline
        1.0.b & -72  & 0   & 30           & -71  & 0   & 37 \\ \hline
        1.1.b & -93  & 0   & 51           & -102 & 0   & 68 \\ \hline
        1.2.b & -103 & 0   & 61           & -108 & 0   & 74 \\ \hline
        1.0.a & -86  & 0   & 44           & -87  & 0   & 53 \\ \hline
        1.1.a & -98  & 0   & 56           & -103 & 0   & 69 \\ \hline
        1.2.a & -112 & 0   & 70           & -102 & 0   & 68 \\ \hline
        1.b   & -120 & 0   & 78           & -129 & 33  & 95 \\ \hline
        O     & -66  & 0   & -            & -69  & 0   & - \\ \hline
        2.0.a & -112 & 0   & 46           & -120 & 0   & 51 \\ \hline
        2.B.a & -    & 100 & -            & -129 & 38  & 60 \\ \hline
        2.0.b & -99  & 0   & 33           & -95  & 0   & 26 \\ \hline
        3.0.b & -124 & 55  & 58           & -114 & 11  & 45 \\ \hline
        3.B.b & -123 & 40  & 57           & -123 & 11  & 54 \\ \hline
        3.B.a & -    & 100 & -            & -128 & 11  & 59 \\ \hline
\end{tabular}
}
\end{center}
\end{table}

\subsection{Experimental evaluation of \mioty{} and \lorawan{} in a private infrastructure}
\begin{figure}[!t]
\begin{center}
\includegraphics[width=0.45\textwidth]{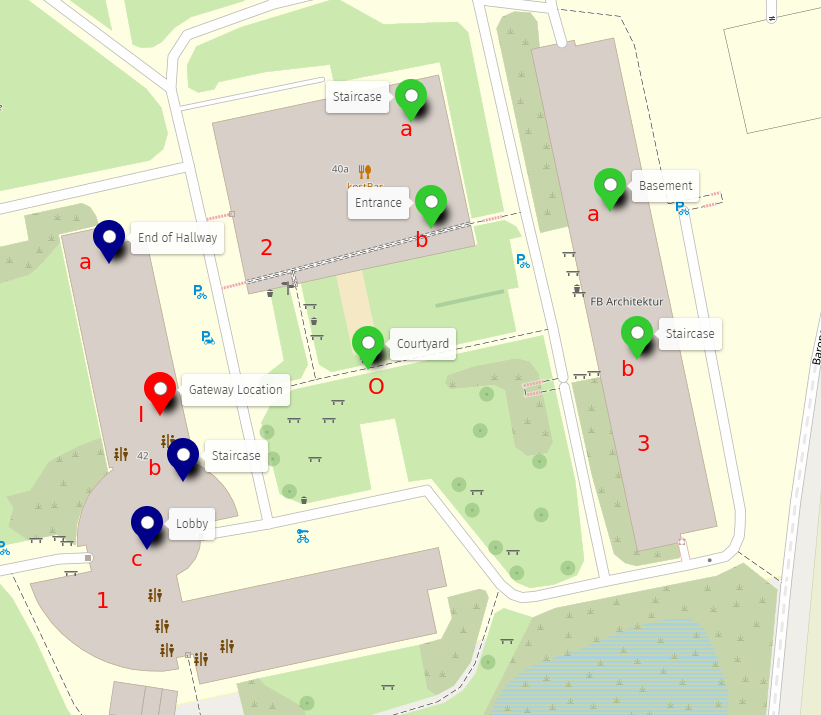}
\end{center}
\caption{Experimental area at campus Emil-Figge-Str. with MPs} %
\label{fig:EFS}
\end{figure}

In this measurement campaign, we compare \mioty{} with \lorawan. 
Since there is currently no publicly available \mioty{} network, a private network was set up for this measurements.
The experiments are performed on university campus Emil-Figge-Str. in Dortmund.
In building 1 the \lorawan{} GW LPS8 and the \mioty{} GW AVA is placed on the first floor near a window (see Fig.~\ref{fig:EFS}).
Two MPs are used as references: MP I is located near the GWs, and MP O is positioned outside, between the three buildings.
The naming scheme for the other MPS is <building>.<level>.<point>, where 'B' denotes the basement.
Fig.~\ref{fig:EFS} shows the experimental area at the campus Emil-Figge-Str. in Dortmund. 
The measured RSSI values at the outdoor MP O correspond to the basic outdoor path loss $L_\mrm{b}$ in (\ref{eq:o2i}).

At each MP, 20 measurements are made for \mioty. For \lorawan, 27 measurements were recorded at each MP, 9 each for SF10, SF11, and SF12.   
The results are shown in Table~\ref{tab:RSSI}.  
The UL RSSI values for \lorawan{} are obtained from the TTN back-end, the UL RSSI values for \mioty{} are obtained directly from the
\mioty{} AVA GW. 
The RSSI values in the Table~\ref{tab:RSSI} are the medians of the measurements at the indicated MP.
The packect error rate (PER) is calculated for \mioty{} and \lorawan{} for all measured values at each MP. 
\lorawan{} can reach the basements in all buildings, \mioty{} can not reach two MPs in basements of building 2 and 3.
The lowest RSSI value measured by the \mioty{} GW is $\unit[-125]{dBm}$, the lowest RSSI value observed by the \lorawan{} GW is $\unit[-138]{dBm}$.   
To compare $L_\mrm{BP}$ of \mioty{} with \lorawan, the RSSI of the MPs is subtracted from the RSSI of the reference MP.
Table~\ref{tab:RSSI} compares $L_\mrm{BP}$ for each indoor MP. 

Due to measurement inaccuracies, these values are not exact. Furthermore, RSSI values are determined using different methods,
which is why it is not possible to rank the performance of the technologies in relation to BPL.
However, the measured values can be used to determine the order of magnitude of BPL to be expected.

\subsection{Long-term evaluation of \mioty{} and \lorawan{}}
In order to evaluate the long-term fluctuation of RSSI values and the PER, two MUNIA-M devices and one \lorawan{} device 
are deployed in building 1 of the experimental environment  (see Fig~\ref{fig:EFS}). 
The device \emph{mioty1} has been installed in a laboratory situated on the third floor. 
The device \emph{mioty2} and the \lorawan{} device have been installed in the server room that is located in the basement.
The GWs have remained in the same location as in the previous experimental evaluation.
The \mioty{} devices send every 15 minutes a message, the \lorawan{} device transmits packets every 15 minutes, each with SF7 through SF12. 
In the course of a month, more than 2,600 messages were sent from each device.
The results of the experiment are displayed in Table~\ref{tab:SNR}. 
The absolute RSSI values are not comparable due to the fact that the \mioty{} devices utilise PCB antennas, 
whereas the \lorawan{} device is equipped with a small rod antenna.
In comparison with the \lorawan{} device, the two \mioty{} devices exhibit a lower PER in relation to the mean RSSI value.
In case of SF12 the \lorawan{} GW provides a better sensitivity and therefore higher MCL compared to \mioty.
The disadvantage of a high SF lies in the longer occupation of the frequency band and higher energy consumption.
\lorawan{} offers a method to automatically optimize data rate, airtime, and energy consumption with Adaptive Data Rate (ADR).
\begin{table}[h!]
\caption{RSSI values in $\unit{dBm}$, SNR in $\unit{dB}$ and PER in \% for \mioty{} and \lorawan{}}
\label{tab:SNR}
    \centering
    \begin{tabular}{|l|r|r|r|r|r|}
    \hline
        RSSI   & max    & median & min    & @SNR  & PER  \\ \hline
        SF7    & -89.0  &  -97.0 & -122.0 &  -9.5 & 12.6 \\ \hline
        SF8    & -90.0  &  -97.0 & -124.0 & -11.8 &  8.8 \\ \hline
        SF9    & -90.0  & -101.0 & -126.0 & -15.2 &  5.1 \\ \hline
        SF10   & -71.0  &  -96.0 & -127.0 & -14.0 &  4.0 \\ \hline
        SF11   & -86.0  &  -95.0 & -130.0 & -16.8 &  4.3 \\ \hline
        SF12   & -64.0  &  -97.0 & -132.0 & -17.2 &  3.1 \\ \hline
        mioty1 & -93.7  & -113.1 & -128.3 &  -2.5 &  1.4 \\ \hline
        mioty2 & -106.5 & -118.9 & -129.2 &  -4.0 & 10.7 \\ \hline
    \end{tabular}
\end{table}%

\begin{figure}[!t]
\begin{center}
\includegraphics[width=0.45\textwidth]{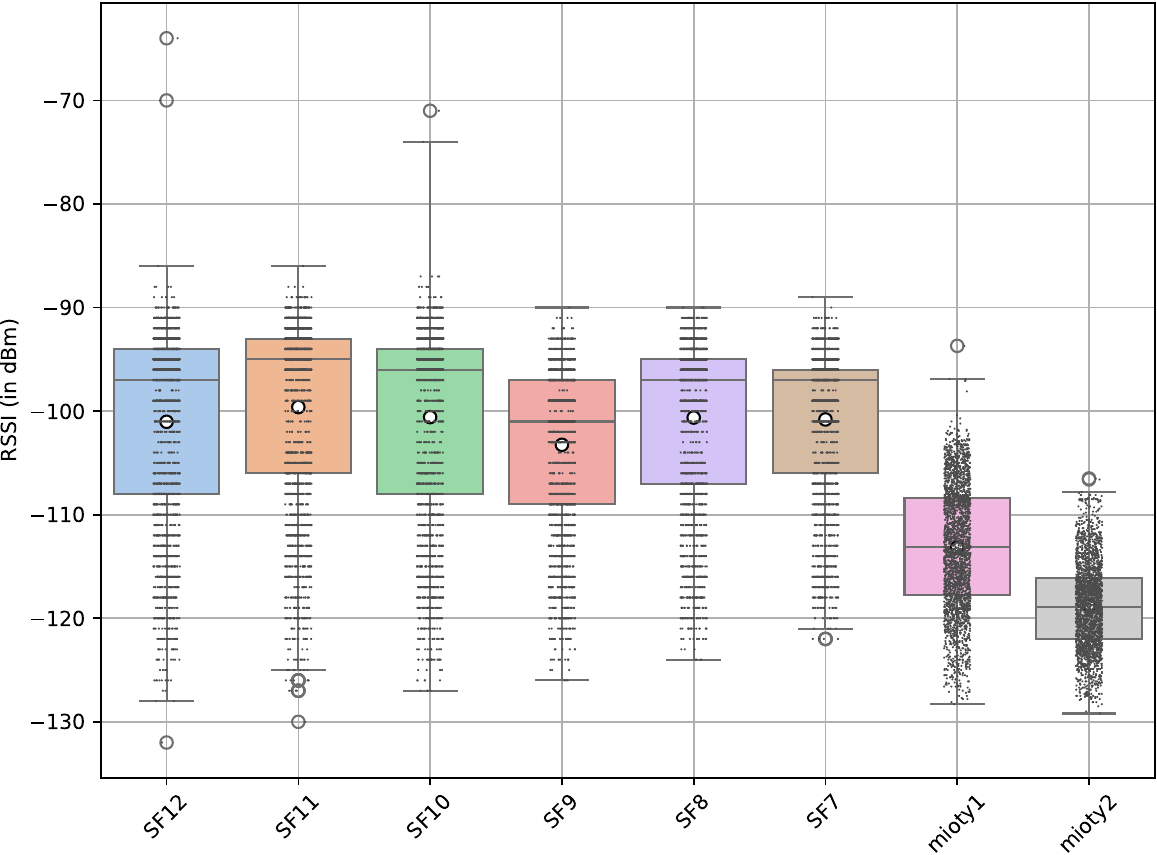}
\end{center}
\caption{Boxplot of measured RSSI values for \mioty{} ans \lorawan} %
\label{fig:box}
\end{figure}

Fig~\ref{fig:box} shows a boxplot of the measure RSSI values. The resolution of RSSI values is \unit[1]{dBm} for \lorawan{}, while it is \unit[0.1]{dBm} for \mioty.
LoRaWAN's RSSI values fluctuate more significantly than those of \mioty, particularly for SF12.

\section{Conclusions}
The paper presents an experimental evaluation of five LPWAN technologies in deep indoor environments. 
NB-IoT, \mioty{} and \lorawan{} show better building penetration than Sigfox and LTE-M.
Because the density of Sigfox BSs is relatively low, indoor coverage depends heavily on the distance to the nearest BSs.
NB-IoT offers higher reliability than other LPWAN technologies by using licensed radio bands and extended coverage levels (ECLs) with repetitions.
Compared to NB-IoT, \mioty{} and \lorawan{} have the advantage of lower power consumption. 
\lorawan{} offers the widest range of energy monitoring devices and the greatest infrastructure flexibility compared to other LPWAN technologies.

\mioty{} is more reliable than \lorawan{} if the RSSI values are high enough. 
In our experiments, the sensitivity of the \mioty{} GW is lower than that of the \lorawan{} GW.
This confirms the specifications listed in the data sheets for the GWs used.
\mioty{} packet loss occured when the average RSSI value drops below $\unit[-120]{dBm}$.
Since the sensitivity of a \mioty{} GW depend mainly on the SDR frontend, this may be a limitation of the \mioty{} GW under test.
\mioty{} is a relatively young technology that is developing rapidly.
Currently, there are fewer providers offering components for \mioty{} compared to \lorawan, 
but \mioty{} may become more important in the future due to its higher reliability and ability to support a higher device density compared to \lorawan. 

The results of the study are integrated in the metering system of the Dortmund University of Applied Sciences and Arts (see \cite{roehrig:iecon25}). 

\bibliographystyle{IEEEtran}
\bibliography{vtc}

\end{document}